\documentclass[conference]{IEEEtran}
\usepackage{graphicx}
\usepackage{mathtext}
\usepackage{amssymb}

\begin{document}

\title{INFLUENCE OF DISTORTIONS OF KEY FRAMES ON VIDEO TRANSFER IN WIRELESS NETWORKS}

\author{\IEEEauthorblockN{E.S. Sagatov, A.M. Sukhov}
\IEEEauthorblockA{Samara State Aerospace University,\\ Moskovskoe sh., 34, \\Samara, 443086, Russia\\
\em e-mails: sagatov@yandex.ru, amskh@yandex.ru}
\and
\IEEEauthorblockN{P. Calyam}
\IEEEauthorblockA{Ohio Supercomputer Center,\\ 1224 Kinnear Road,\\ Columbus, OH 43212, USA \\
\em e-mail: pcalyam@osc.edu}}

\maketitle

\begin{abstract}
In this paper it is shown that for substantial increase of video quality in wireless network it is necessary to execute two obligatory points on modernization of the communication scheme. The player on the received part should throw back automatically duplicated RTP packets, server of streaming video should duplicate the packets containing the information of key frames. Coefficients of the mathematical model describing video quality in wireless network have been found for WiFi and 3G standards and codecs MPEG-2 and MPEG-4 (DivX). The special experimental technique which has allowed collecting and processing the data has been developed for calculation of values of factors.
\end{abstract}

{\bf Categories:}\\
{H.4.3} {\bf [Communications Applications]} {Computer conferencing, teleconferencing, and videoconferencing}\\
{C.2.1} {\bf [Network Architecture and Design]} {Wireless communication}\\
{C.2.5} {\bf [COMPUTER-COMMUNICATION NETWORKS]} {Local and Wide-Area Networks} {Internet (e.g., TCP/IP)}


\section{Introduction}
\label{intr}

Mobile solutions in the telecommunications area are being increasingly adopted in everyday life. Modern cellular telephones possess the same capabilities of a traditional computer. Laptops and netbooks provide a high degree of mobility. Owing to their compactness and presence of 3G, WiFi, and WiMAX network adapters, we are experiencing pervasive communications as and when needed. According to data Cisco VNI~\cite{cisco}, the Internet traffic on wireless devices has annual growth of more than 250\%. By 2013 the volume of such traffic will increase by 66 times in comparison with 2008 and will make up 5,4\% of all IP-traffic in the Internet. By 2013 various video content will make 64\% of all traffic of wireless networks in the world.

There is serious obstacle for daily use of video services in wireless networks: the corresponding communication quality is not satisfactory. High degradation levels of packet losses and delay variation (network jitter) in wireless networks affect data transmission in real time modes of communication such as the multimedia traffic, and make them susceptible to distortions in user experience. Packets of streaming video experience drastic changes in sequence numbers or are lost during transfers on network paths because of considerable delay variation. At the receiver-side video image, distortions manifest as plural artifacts, lack of lip synchronization, and sometimes frame freezing.

The technique of certification of packet networks in recommendation RFC-2544~\cite{rfc2544} defines the following key parameters of network quality: available bandwidth, packet delay, network jitter, number of the lost packets, quantity of packages with errors.

The idea of subjective testing ({\it MOS}) consists that video received after transfer on network, is shown to a commission of experts who put down scores, being based on the impressions of quality. The initial clip is coded by one of codecs MPEG-2, MPEG-4 or Windows Media Video 9 and is passed through a wireless network supporting standards such as WiFi, 3G or WiMAX. There are many methods of demonstration of sequences and gathering of the estimations, some of them are described in recommendations ITU~\cite{itu}. Unfortunately, they are calculated based on video comparison in broadcasting format, and aren't so convenient for testing carrying out on PCs.

In the present paper the problem of adaptation of modern coding algorithms and transfer of video for wireless networks, such as 3G, WiFi and WiMAX, and also for other networks with distorted quality characteristics is considered. In papers~\cite{cs,ct} it has been shown that subjective scope of quality of video as well as network parameters, has gradation Good, Acceptable and Poor (GAP). Such description helps to understand the qualitative dependences arising at network translation of video and to make primary numerical approach.

In the present paper, we attempt to find numerical dependence of video quality on network parameters. Distinctive feature of our approach is that the specified dependence is described by a simple mathematical model that allows us to compare numerical values of coefficients. On the basis of similar comparison, we find not only the most essential factors influencing quality of video, but also compare them for various codecs.

In paper, another contribution is the characterization of distortions which damage the "key" frames as well as the "usual" frames is considered. Key frame is the frame which bears in itself the full information on the video image and it can be restored without reinforcement with the additional data. Usual frame is one that codes difference between the previous frame and current one. Degree of key frame compression is less than at usual shot, and the size several times is more. In this paper numerical comparison of influence of the errors which have occurred on key frames, for codecs MPEG-2 and MPEG-4  are given.

This paper is organized as follows: Section 1 describes about premises for mathematical modeling of dependence of video quality from characteristics of network connection, in Section 2 planning of experiments is explained, and results of processing of experiments are presented in Section 3, Section 4 tells about numerical results of experiment and parameters of mathematical model.

\section{Premises for modeling}
\label{sb}

The communication quality of video transmission worsens depending on characteristics of network connection. In order to describe quality of transferred video depending on network parameters the universal function $Q_{MOS}(p,j,D,B)$, describing video quality on a {\it MOS} scale has been considered. This function can be expanded in a power series on network parameters, thus it is possible to be limited to linear members.

For the fixed speeds of video stream it is enough to consider only a linear dependence from two parameters (losses of packets and network jitter)~\cite{scdi}:
\begin{equation}
Q_{MOS}=Q_{ideal}-\alpha p- \beta j,
\end{equation}
 	where 	 	
\begin{itemize}
	\item 
	$Q_{ideal}$ - maximum quality of video for the given codec, points from zero to five;
	\item 
	$p$ - packet loss, \%;
	\item 
	$j$ - network jitter (delay variation), {\it sec};
	\item 
  $Q_{MOS}$ - video quality on received side, points from zero to five;
	\item 
	$\alpha,\beta$ - coefficients of model which can be found experimentally.
\end{itemize}

The uniform video sequence which was compressed by codecs MPEG-4 (DivX), MPEG-2 and Windows Media video 9 with constant bitrate 256 Kbps has been developed for experimental tests.

The basic aim  put in the present research is revealing of influence of key frames on quality of received video. Therefore coefficients $\alpha^k$ and $\beta^k$ for a stream with damage of key frames, and also $\alpha^w$ and $\beta^w$ for sequences without damage of key frames are separately calculated. For example, $\alpha_{DivX}^k$ will designate coefficient which characterizes deterioration of video encoded by MPEG-4 (DivX) if a key frame was damaged. Further in paper coefficients $\alpha_{DivX}^k$ and $\alpha_{DivX}^w$, as well as $\alpha_{Mpeg2}^k$ è $\alpha_{Mpeg2}^w$, will be found and compared to define in numerical sort a measure of video deterioration.

\section{Experiment planning}
\label{s2}

For finding of coefficients from the Equation (1) we had been developed and spent a number of experiments. Videos files encoded by MPEG-4 (DivX), MPEG-2 and Windows Media Video 9 were transferred on a notebook connected to wireless network WiFi, WiMAX or 3G. On a notebook record of received video was spent to a file, the network traffic at packet level was in a parallel way written by network sniffer Wireshark. Thus, under the received video it is possible to find video quality on scale MOS, and parameters of network connection may be calculated on network logs.

\begin{figure*}
\centering
\includegraphics[height=5cm]{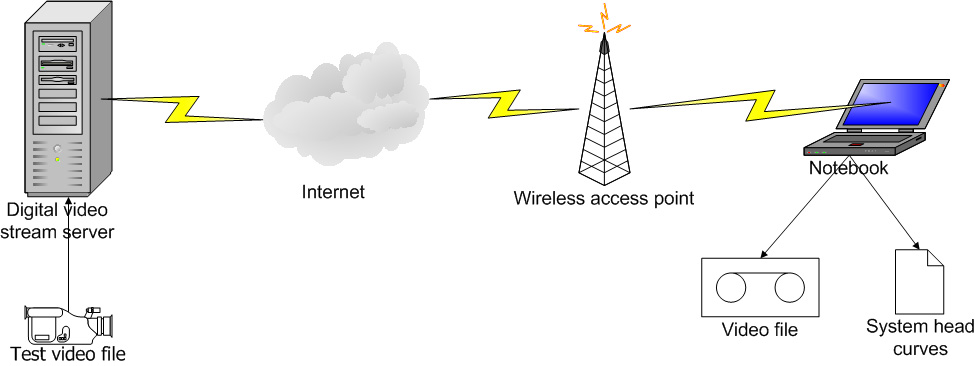}
\caption{Scheme of experiment}
\label{f1a}
\end{figure*}

The software was applied to carrying out and the analysis of experiments (see Fig.~\ref{f1a}):

\begin{enumerate}
	\item 
VLC media player~\cite{vlc} was used as a program video server and a video player with possibility of record of video received on a network in a file on a notebook.
	\item 
The Wireshark Network Protocol Analyzer~\cite{wire}, with its help on a notebook all network traffic was written, and then analyzed.
	\item 
VirtualDub~\cite{virt}, by means of this program the frame analysis of the video was made for calculation MOS.	
		\item 
AviSynth 2.5~\cite{avis}, with its help video encoded WMV was parsed in program VirtualDub. The given codec works using technology DirectShow, instead of VFW (Video For Windows), and cannot be opened directly in VirtualDub.		
\end{enumerate}

All video fragments written during experiments and network logs are published on site of Internet TV ltd~\cite{trace}. 

Experiments will look equally for networks Wi-Fi, WiMAX and 3G, only network equipment varies. In the Figure~\ref{f1a} the scheme of experiment is presented.

Alone video clip with various types of the picture has been prepared for experiment: static, with weak movement, with fast movement, with brightness change. Then this video clip has been encoded with usage of MPEG-4 (DivX), MPEG-2 and Windows Media Video 9. Following parameters of video are installed:
\begin{itemize}
	\item 
Video Resolution 320 x 240 {\it pixels}
	\item 
Frame Rate - 24 {\it Frames per second}
	\item 
Bitrate - 256 {\it Kbps}
	\item 
Maximal quality
\end{itemize}

Files of initial video are also published on site of Internet TV ltd~\cite{trace}.

Software product Videolan~\cite{vlc} was used as the program for network video translation. This program was started from a server connected to an Internet with real IP address.

VLC media player and network sniffer Wireshark writing  network traffic have been installed on the computer which accepted video. VLC media player has been customized on simultaneous demonstration of accepted video on the screen and its saving in a video file. Thus, at carrying out of each experiment two files for the further analysis are written: a video file and a file of network logs.

For carrying out of experiments we were used LAN of Samara State Aerospace University (WiFi), the Samara segments of all-Russian operators: the Megaphone (3G), a Beeline (3G) and Metromaks (WiMAX) which have free of charge given the equipment and technical possibilities for test trials.

\section{Processing of results of experiments}
\label{s4}

\begin{figure*}
\centering
\includegraphics[height=6cm]{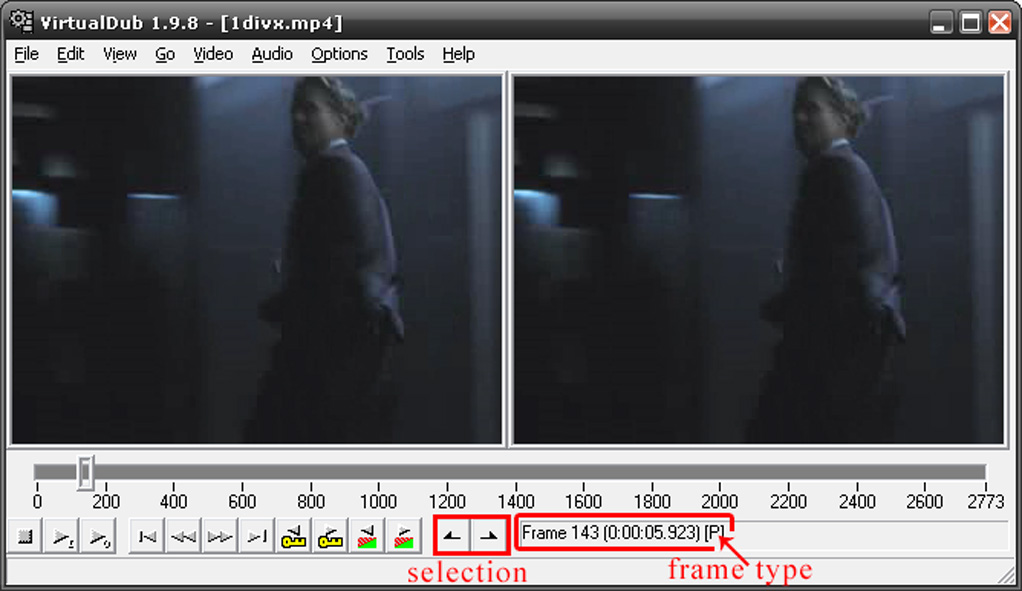}
\caption{Search of distortion frames}
\label{f2}
\end{figure*}

\begin{figure*}
\centering
\includegraphics[height=9cm]{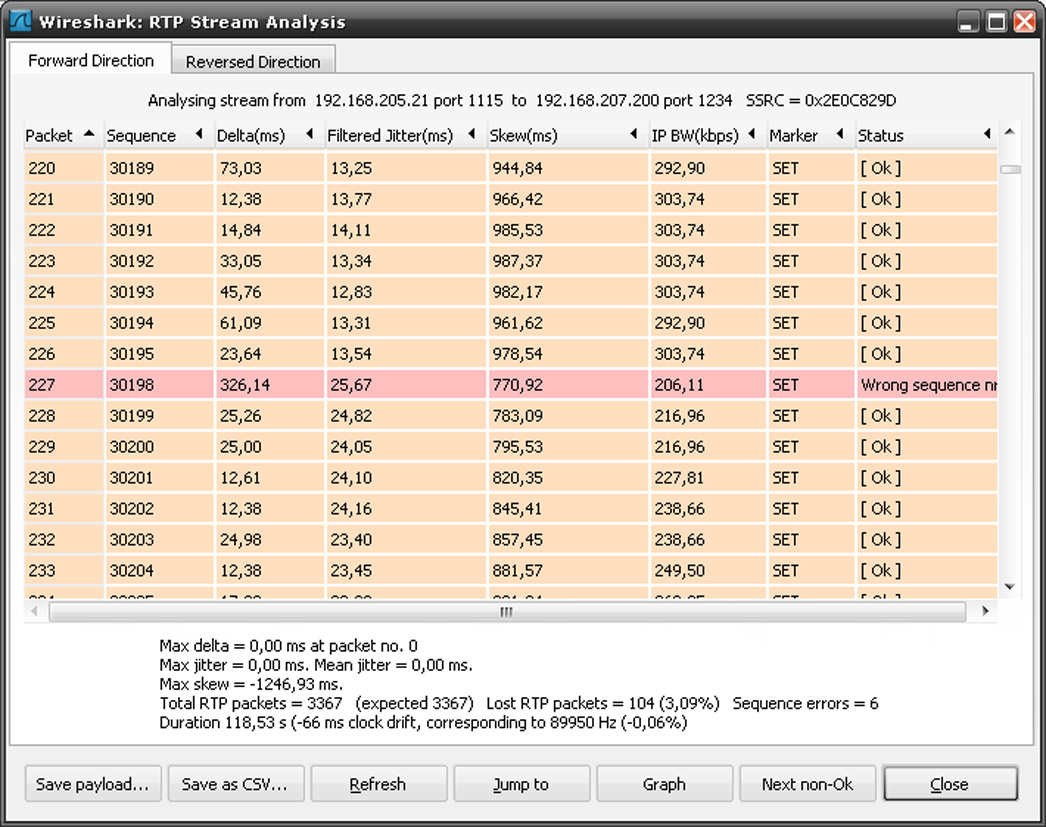}
\caption{Error of RTP packet sequence}
\label{f3}
\end{figure*}

\begin{figure*}
\centering
\includegraphics[height=9cm]{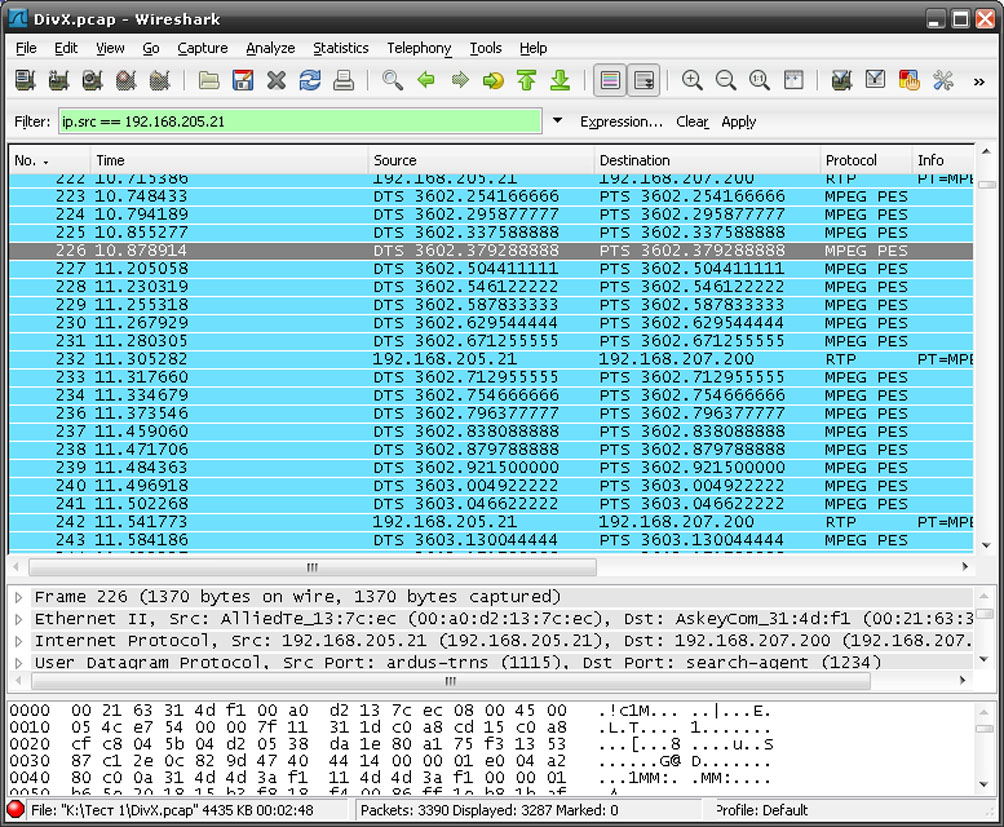}
\caption{Network logs}
\label{f4}
\end{figure*}

For a network of Wi-Fi standard periodic deterioration of network, accordingly the picture is characteristic worsens during the corresponding moments. The order of the analysis of all received video fragments and network logs is identical. Further process of the analysis of video files and a corresponding network logs during the moments of video deterioration is described:
\begin{enumerate}
	\item 
The Program VirtualDub opens the recorded video file. The frame previous the distorted frame is searched, in the Figure~\ref{f2} it is a frame 143 which is displayed on 5923's millisecond {\it ms} of a video series. Similarly a last distorted frame is searched, in a described case it is 171's frame shown on 7083 millisecond. Thus, duration of distortion consists of 28 frames or 1160 {\it ms}.	
	\item 
In the network sniffer Wireshark the corresponding logs (http://stream.ip4tv.ru/wireless/WiFi/test1/1divx.pcap) are loaded. For RTP packets Wireshark has the built in analyzer, the necessary stream is selected from it for more detailed analysis. In the analyzer (the Figure~\ref{f3}) is resulted the list of all packets and red places where packages have been lost are flagged. Also the statistical data on an inter-packet interval and jitter is specified. From column Sequence it is visible that are accepted a package with number 30195 and a package with number 30198, and between them two packages have been lost.
	\item 
The main problem of the analysis - how to correlate a file of network logs with video recording. For this purpose in a network logs packet (226, see Figure~\ref{f4}) is searched after which there was an error. In relation to the beginning of record of dens this package has arrived on 10,87 second. In item 1 it has been found that in a video distortion has occurred on 143 frame which was displayed on 5923 {\it ms} of a video series. Thus, the required relation is defined.
	\item 
For reception of authentic statistical calculations it was accepted that the length of sequence for the analysis should be multiple to 100 packets. Now it is not difficult to count up percent of the lost packages.
	\item 
In Analyzer RTP the inter-packet interval and network jitter also is specified.
\end{enumerate}

Other major task arising at processing of experimental results is an estimation of quality of video on scale MOS. Algorithm of estimation the following:
\begin{enumerate}
	\item 
Labels in program VirtualDub (the Figure~\ref{f2}) are installed on the first distorted frame of video and last one. 	
	\item 
The distorted video can be viewed some times and to compare it to the original. 	
	\item 
Quality of the video at the moment of an error on scale MOS from 1 to 5 is visually estimated.	
\end{enumerate}

In the present paper for each error the estimation was exposed by 4 experts. Then by their estimations arithmetic mean value was calculated.

At processing of the files encoded by WMV9 codec there were complexities. Unfortunately, program VirtualDub does not display key and not key frames for video encoded by the codec of Windows Media Video 9, therefore experiments with it us have not been handled.

\section{Analyzing of results of experiments}
\label{s5}

Obtained data has been handled by the technique described in the previous section. All errors, both at video level, and on a network layer have been parsed. Subjective quality of video $Q_{MOS}$ depending on packet loss $p$ and network jitter $j$ is found. The received values of coefficients are gathered in Tables~\ref{t1} and~\ref{t2}.

\begin{table*}
	\centering
		\begin{tabular}{|c|c|c|c|c|c|c|} \hline
$N$ & Codec &	$Q_{ideal}$ &	$\alpha^k$ & $\beta^k$  &	 $\alpha^w$ & $\beta^w$\\ \hline
1 &	MPEG-2 & 4.2$\pm$0.2 &	0.15$\pm$0.03	& 0.011$\pm$0.002 &	0.04$\pm$0.01 &	0.003$\pm$0.001 \\
2 & DivX  & 4.7$\pm$0.2 &	0.27$\pm$0.05	& 0.013$\pm$0.003 &	0.13$\pm$0.02 &	0.01$\pm$0.002 \\ \hline
	\end{tabular}
	\caption{Values of coefficients of model for codecs MPEG-2, DivX in WiFi networks}
	\label{t1}
\end{table*}

\begin{table*}
	\centering
		\begin{tabular}{|c|c|c|c|c|c|c|} \hline
$N$ & Codec &	$Q_{ideal}$ &	$\alpha^k$ & $\beta^k$  &	 $\alpha^w$ & $\beta^w$\\ \hline
1 &	MPEG-2 & 4.2$\pm$0.2 &	0.005$\pm$0.002	& 0.005$\pm$0.002 &	0.004$\pm$0.001 &	0.003$\pm$0.001 \\
2 & DivX  & 4.7$\pm$0.2 &	0.01$\pm$0.003	& 0.003$\pm$0.001 &	0.002$\pm$0.0005 &	0.002$\pm$0.0008 \\ \hline
	\end{tabular}
	\caption{Values of coefficients of model for codecs MPEG-2, DivX in 3G networks}
	\label{t2}
\end{table*}

It has been found that quality of the video depends not only on percent of packet losses and network jitter, but also from frame type on which there was an error. Key frame  carries in itself the complete information on the video  and can be restored without engaging of the additional data, and usual one is a frame which encodes a difference between the previous frame and flowing. Accordingly, if the error damages a key frame quality of the video worsens more strongly in comparison with a similar error in a usual frame. Therefore in Tables~\ref{t1},~\ref{t2} it is specially selected two types of coefficients - with losses on key frames and without them.

In Tables~\ref{t1},~\ref{t2} $\alpha^k$ and $\beta^k$  are coefficients of model with losses of packets on key frames, $\alpha^w$ and $\beta^w$ are coefficients for the intact key frames, and $Q_{ideal}$ is an estimation on scale {\it MOS} for a source file (before transfer on a network).

Initially for the video encoded by MPEG-4 (DivX), quality is above, than for video encoded MPEG-2, but at deterioration of characteristics of a network quality decreases considerable and at the big network interferences becomes similar to quality MPEG-2. In case of damage of a key frame video quality  encoded MPEG-4 (DivX) will fall more than in 2 times at the same characteristics of a network, and MPEG-2 in 3.75 times.

Thus, for substantial increase of video quality by transmission to a wireless network it is necessary to fulfill three mandatory items on upgrade of the circuit of link:
\begin{enumerate}
	\item 
The server of stream video should duplicate the packets containing the information of key frames	
	\item 
To upgrade a player on the receiving side automatically to throw back duplicated RTP packets
	\item 
The period between key frames cannot exceed 2 seconds (optimally 1 second)	
\end{enumerate}

These simple measures will lead to what even poor (GAP)~\cite{cs} networks will translate video with estimation more then 3.5. It is necessary to notice that for codecs MPEG-4 (DivX) and MPEG-2 the transferred size of the traffic will increase on 7-10\%, and quality of video grows not less than in 2 times. Some researched video players at playback MPEG-TS MPEG-4  streams already automatically discard repeating packets.

The experiments spent on network WiMAX, have shown very good stability of the given network to link deterioration. WiMAX networks under the characteristics are comparable to fixed networks Ethernet. It is very difficult to find network errors since the percent of losses for all spent tests is near 0\%, and a variation of delay of an order 20 msec even in the tests spent at the big competing traffic. According to paper~\cite{cs} such type of the traffic on system GAP concerns to good (Good).

3G networks are very sensitive to external interferences and even at good level of a signal there are considerable losses of packets (Poor) and a delay variation of an order 40 msec (Acceptable). Also it has been found that for one of 3G providers the equipment frequently duplicated outgoing packages that created on video very strong artefacts. New player is needed for eliminating the given interferences.

\section{Conclusion}
\label{s7}

There is an increasing demand of voice and video applications on wireless devices due to the recent developments in smart phones and tablet PCs. To cater to this demand efficiently and reliably, it is important for the application developers and service providers to characterize and tune the performance of RTP streams that deliver the content. In addition, the area of video performance measurement is in its early stages, and developing effective techniques to measure video quality is vital. In this paper, we address both the above dimensions of requirements. We show scenarios where RTP packets are duplicated and evaluate how the video player should handle these duplicated packets. We also show a novel experimental technique to identify key frames in video, and evaluate how redundancy of those key frame packets can improve video quality. 

A significant contribution of our work is the development of a mathematical model for estimating video quality of codecs MPEG-2 and MPEG-4 (DivX) codecs used in open-source VLC on WiFi and 3G standards compliant wireless networks. Our model development has been done with due considerations given to the handling of duplicate packets, and adding redundancy to key video frames. 

The video quality measurement experimental technique and research findings presented in this paper have been implemented in our Internet video broadcasting efforts at SSAU, Togliatty branch SSAU and Internet TV service.
Owing to these implementations, we have observed that our Internet video broadcasting service offering has become more automated, predictable and our service has seen notably lowered operation costs.

In summary we would like to thank Leonid Fridman, the Professor from University of Mexico for fruitful dialogues in which course the idea of this article has taken shape. Also it would be desirable to thank all collective of technical service RIPE ncc and especially Ruben van Staveren and Roman Kalyakin for constant assistance in comprehension of subtleties of a measuring infrastructure. We also would like to express the gratitude to the Wolfram Research corporation, which the first has marked our preprint and has given us licenses for the right of use of 
{\it Mathematica}.

\end{document}